# The impact of sunlight on high-latitude equivalent currents

K. M. Laundal[1,2], J. W. Gjerloev[1,3], N. Østgaard[1], J. P. Reistad[1], S. Haaland[1,4], K. Snekvik[1], P. Tenfjord[1], S. Ohtani[3], and S. E. Milan[1,5]

[1]Birkeland Centre for Space Science, University of Bergen, Bergen, Norway, [2]Teknova AS, Kristiansand, Norway, [3]The Johns Hopkins University Applied Physics Laboratory, Laurel, Maryland, USA, [4]Max Planck Institute for Solar Systems Research, Göttingen, Germany, [5]Department of Physics and Astronomy, University of Leicester, Leicester, UK

**Key Points:**

- Global equivalent currents are investigated with respect to the position of the sunlight terminator
- The global equivalent current resembles Hall currents on the dayside but not on the nightside
- The day-night contrast increases with solar EUV flux

**Correspondence to:**
K. M. Laundal,
karl.laundal@ift.uib.no

**Abstract** Ground magnetic field measurements can be mathematically related to an overhead ionospheric equivalent current. In this study we look in detail at how the global equivalent current, calculated using more than 30 years of SuperMAG magnetometer data, changes with sunlight conditions. The calculations are done using spherical harmonic analysis in quasi-dipole coordinates, a technique which leads to improved accuracy compared to previous studies. Sorting the data according to the location of the sunlight terminator and orientation of the interplanetary magnetic field (IMF), we find that the equivalent current resembles ionospheric convection patterns on the sunlit side of the terminator but not on the dark side. On the dark side, with southward IMF, the current is strongly dominated by a dawn cell and the current across the polar cap has a strong dawnward component. The contrast between the sunlit and dark side increases with increasing values of the $F_{10.7}$ index, showing that increasing solar EUV flux changes not only the magnitude but also the morphology of the equivalent current system. The results are consistent with a recent study showing that Birkeland currents indirectly determine the equivalent current in darkness and that Hall currents dominate in sunlight. This has implication for the interpretation of ground magnetic field measurements and suggests that the magnetic disturbances at conjugate points will be asymmetrical when the solar illumination is different.

## 1. Introduction

It has been known for three centuries that geomagnetic disturbances correlate with auroral activity [see, e.g., *Egeland and Burke*, 2010]. More than 100 years ago, Kristian Birkeland produced maps of such disturbances, in terms of an overhead equivalent current, which indicated that they tend to appear in a two-cell structure whose orientation is fixed with respect to the Sun [*Birkeland*, 1908]. We now understand this pattern in terms of the Dungey cycle [*Dungey*, 1961] plasma circulation in the magnetosphere-ionosphere system, which is associated with horizontal Hall currents that by definition are antiparallel to the convection.

Hall currents are not the only currents that flow in the ionosphere. A standard way of decomposing the ionospheric currents is to consider the height-integrated horizontal currents and magnetic field-aligned currents separately. The former can be decomposed further relative to the convection electric field: Hall currents flow in the $\mathbf{B}\times\mathbf{E}$ direction, and Pedersen currents flow in the direction of $\mathbf{E}$. Relating the equivalent current to these current components is not straightforward, however, and depends on the ionospheric conductivity.

Elaborate investigations have shown that the equivalent current on average is slightly skewed with respect to the Sun [e.g., *Vestine et al.*, 1947], with the current across the polar cap pointing slightly dawnward. The statistical study by *Friis-Christensen and Wilhjelm* [1975] clearly showed that the equivalent currents are more skewed during winter than in summer. The rotated global patterns reported in these studies are not consistent with the expected direction of the Hall current, according to several statistical studies of ionospheric convection [e.g., *Heppner and Maynard*, 1987; *Weimer*, 2005; *Haaland et al.*, 2007; *Pettigrew et al.*, 2010]. The deviation between the Hall current system and the equivalent current was explained by *Vasyliunas* [1970], based on the Fukushima theorem [e.g., *Vasyliunas*, 2007; *Fukushima*, 1994, and references therein]. *Vasyliunas* [1970] argued that the equivalent current is the divergence-free part of the total height-integrated horizontal current, which not necessarily coincides with the Hall current. The equivalent current is equal to the Hall current only when the conductance is uniform or the conductance gradients are perpendicular to convection streamlines [*Laundal et al.*, 2015].

In a recent study *Laundal et al.* [2015] looked at simultaneous measurements of the magnetic field on ground and in space during different sunlight conditions. They particularly focused on polar cap latitudes, poleward of the auroral oval. They found that in this region the equivalent current is typically antiparallel to the curl-free current in darkness. Since the total current can be written as the sum of a divergence-free field and a curl-free field, this finding is consistent with the actual current being close to zero, so that its divergence-free and curl-free components balance. If this is true, ground magnetic field measurements at high latitudes depend on the surrounding Birkeland (field-aligned) current system in darkness. In sunlight, the equivalent currents were found to be largely antiparallel to simultaneous measurements of the convection, indicating that it is dominated by Hall currents.

The study by *Laundal et al.* [2015] focused on very high latitudes and primarily point-by-point comparisons between ground magnetic field perturbations and measurements from space. In this study, which to a great extent builds on the results by *Laundal et al.* [2015], we perform a detailed investigation of the global equivalent current pattern at high latitudes and investigate its morphology with respect to a precisely determined sunlight terminator. We also investigate the effect of solar EUV flux, parametrized by the $F_{10.7}$ index. The current patterns are calculated using a classical technique based on spherical harmonic analysis, described by *Chapman and Bartels* [1940]. We do the calculations in the nonorthogonal quasi-dipole coordinate system [*Richmond*, 1995], such that variations due to nondipole features in the Earth's magnetic field are reduced. The technique is described in detail in the next section. The equivalent current patterns are presented in section 3 and discussed in section 4. Section 5 concludes the paper.

## 2. Technique

In this section we present the details of the magnetometer data selection and processing.

### 2.1. Calculation of Equivalent Currents

The basis for this study is data from ground magnetometers, processed and provided by the SuperMAG collaboration [*Gjerloev*, 2009, 2012]. SuperMAG provides the data in a common format and coordinate system. Baselines are also subtracted using a common method across all magnetometers. The baseline removal also removes recurrent diurnal variations, primarily associated with solar quiet currents. Details about the data processing can be found in *Gjerloev* [2012]. The coordinate system used by SuperMAG is a local magnetic system, in which the $N$ component is defined to lie along the measured dominating horizontal direction, determined in 17 days sliding windows. $Z$ points down, and $E$ completes a right-handed system ($E = Z \times N$).

The ground magnetometer data are used to calculate ionospheric equivalent currents. To this end, four data processing steps are used.

#### 2.1.1. Step 1: Conversion to Quasi-Dipole Coordinates

We use the apex quasi-dipole (QD) coordinate system [*Richmond*, 1995], defined in terms of field line tracing along magnetic field lines in the International Geomagnetic Reference Field (IGRF). The three QD coordinates are longitude, latitude, and geodetic height ($\phi_q$, $\lambda_q$, and $h$, respectively). Using this coordinate system, it is possible to reduce the effect of longitudinal variations in the terrestrial magnetic field and present the data approximately as if the field was a dipole. Because of the nondipole terms in the IGRF, the definition of QD coordinates implies that the resulting coordinate system is nonorthogonal. That also means that the vector components should be scaled to take into account the contraction and expansion of the coordinate grid. We follow *Richmond* [1995] and calculate the vector components as follows:

$$\begin{aligned} B_{\phi_q} &= \frac{\mathbf{f}_1 \cdot \mathbf{B}}{F} \\ B_{\lambda_q} &= \frac{\mathbf{f}_2 \cdot \mathbf{B}}{F} \\ B_h &= \frac{\mathbf{k} \cdot \mathbf{B}}{\sqrt{F}}, \end{aligned} \tag{1}$$

where $\mathbf{f}_1$ and $\mathbf{f}_2$ are QD base vectors defined by *Richmond* [1995] and $F = \mathbf{f}_1 \times \mathbf{f}_2 \cdot \mathbf{k}$ and $\mathbf{k}$ is an upward unit vector. The base vectors are available through software published by *Emmert et al.* [2010] and are defined in terms of geodetic components. In order to recover the geodetic magnetic field components from those provided by SuperMAG, we make the assumption that the SuperMAG $N$ component represents the field in the direction of the horizontal component of the IGRF. Thus, we use the IGRF declination angle at each magnetometer station (taking secular variations into account) to rotate the vectors to geodetic coordinates. The validity of this method was investigated by *Laundal and Gjerloev* [2014], and the mean error was found to be 0.4° in a sample of 106 observatories with well-determined alignment. However, stations which are located at local magnetic anomalies, which are not captured by the IGRF, will result in some outliers.

The components in equation (1) correspond to the following decomposition of the magnetic field:

$$\mathbf{B} = B_{\phi_q}\mathbf{f}_2 \times \mathbf{k} + B_{\lambda_q}\mathbf{k} \times \mathbf{f}_1 + \sqrt{F}B_h\mathbf{k}. \qquad (2)$$

$B_{\phi_q}$ and $B_{\lambda_q}$ can be interpreted as the magnetic field projected on horizontal unit vectors pointing perpendicular to QD meridians and circles of latitude, scaled by the geographic length per QD length along these vectors. The vertical component, $B_h$, is assumed to scale inversely with the linear dimension of the current system. This means that the components can be interpreted in terms of currents that are ordered in the QD coordinate system. Consequently, effects of local variations in the Earth's magnetic field are reduced [*Gasda and Richmond*, 1998; *Laundal and Gjerloev*, 2014] and so are the effects of secular variations. Because the field varies, the base vectors and magnetometer locations also vary slowly.

It is implicit in the assumption of invariance with respect to nondipole field structures that the components represent the magnetic perturbations in a dipole field on a spherical Earth. In that case, the base vectors would be unit vectors, and the components would be the same as in a geographic system. Therefore, we also interpret the estimated equivalent currents as representative of a dipole field and spherical Earth.

#### 2.1.2. Step 2: Spatial Binning

We have used data from all SuperMAG magnetometers at ≥49° latitude (all coordinates are in QD). The data have then been sorted according to an equal area grid in magnetic local time/MLAT, with 920 bins. The bins are spaced by 2° latitude and extend from 49° to 89°. The [87°, 89°] latitude sector is separated in $M_0 = 8$ different sectors. The following sectors must then be divided in $M_{i+1} = M_i \frac{3+i}{2+i}$ sectors for the cells to have the same area in polar coordinates. The grid is shown in Figure 1a. When more than one magnetometer is within a certain bin, we calculate the mean of each component. In this way, we end up with a time series from each grid cell, with gaps whenever there are no magnetometer present in the cell. These time series are the basis for the further analysis.

#### 2.1.3. Step 3: Temporal Binning and Computation of Average Vectors

Depending on the question at hand, we group the data in the bin time series according to external conditions and calculate the mean. This gives up to 920 average vectors on the polar coordinate system. These vectors are used in the inversion in the next step.

Figure 1b shows the averages of the horizontal components of the vectors in each bin. These calculations are based on data from 1980 to 2014. The vectors have been rotated 90° clockwise to align with an overhead line current. The total number of samples used to make this plot is more than 1.1 billion. Later plots will be based on data from 1981 to 2014, since the OMNI database does not include 1980. OMNI provides measurements of the solar wind and the interplanetary magnetic field at 1 min cadence whenever available, time shifted to the bow shock.

#### 2.1.4. Step 4: Estimate Magnetic Potentials and Compute Current Function

The binned average magnetic field measurements are finally used to fit a spherical harmonic representation of a magnetic scalar potential. The scalar potential is associated with what is assumed to be a static magnetic disturbance field, representative of the external conditions by which the data are constrained. The ionospheric equivalent current is then calculated from the magnetic potential. The method we use for this is basically the same as the classical technique described by *Chapman and Bartels* [1940] and used frequently since then [e.g., *Vestine et al.*, 1947; *Kamide et al.*, 1981; *Friis-Christensen et al.*, 1984].

The field components defined in equation (1) can, according to *Richmond* [1995], be approximately related to an equivalent current in QD coordinates by treating the QD coordinates as orthogonal spherical coordinates. Then $\mathbf{B} = -\nabla V$ and $\nabla^2 V = 0$ with $\nabla$ and $\nabla^2$ in the standard spherical form. The spherical harmonic representation of the magnetic potential $V$ is then

$$V(r, \lambda_q, \phi_{MLT}) = a \sum_{n}^{30} \sum_{m}^{\max(n,10)} P_n^m(\lambda_q) \left[ \left( A_{n,e}^m \left(\frac{r}{a}\right)^n + A_{n,i}^m \left(\frac{a}{r}\right)^{n+1} \right) \cos(m\phi_{MLT}) + \left( B_{n,e}^m \left(\frac{r}{a}\right)^n + B_{n,i}^m \left(\frac{a}{r}\right)^{n+1} \right) \sin(m\phi_{MLT}) \right], \qquad (3)$$

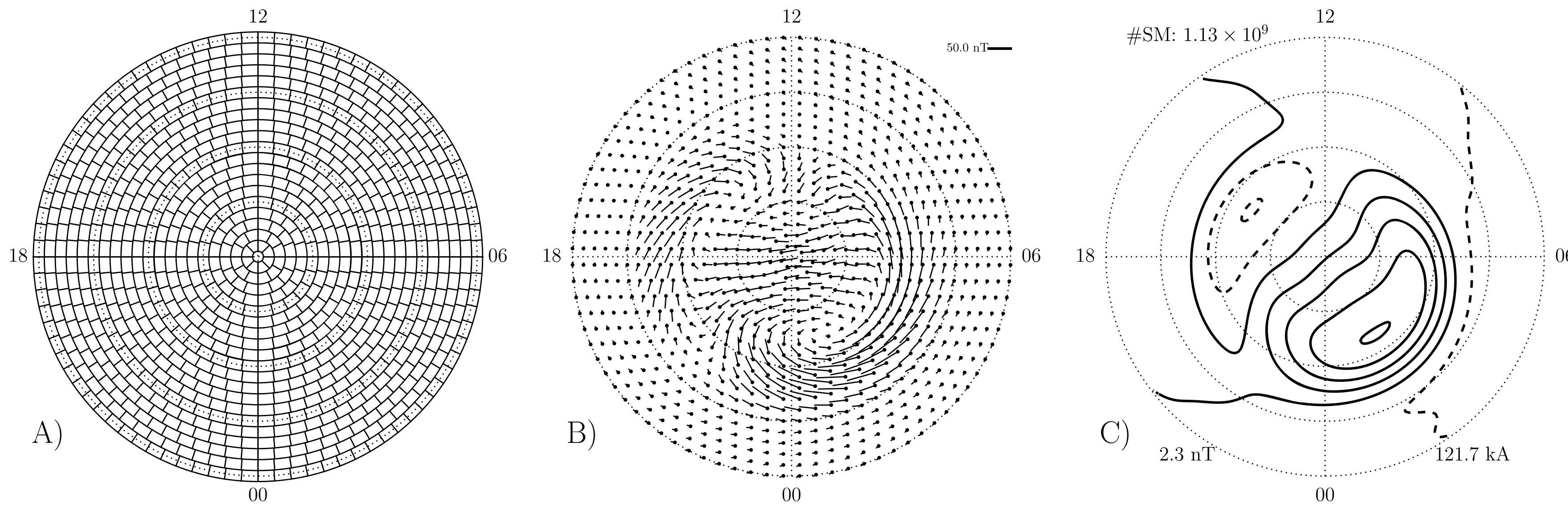

**Figure 1.** Illustration of data processing done to arrive at equivalent current functions. (a) The equal area grid which is used, (b) average horizontal magnetic field vectors, rotated 90° to align with an overhead current, and (c) the equivalent current based on the external part of the magnetic potential. This example is based on all available SuperMAG data at >49° between 1980 and 2014. The total number of data points on which the current is based is shown in the top left corner, in this case more than 1.1 billion. The lower left corner shows the weighted root-mean-square difference between the fitted magnetic field and the average vector components. The lower right corner shows the total current flowing between the maximum and minimum of the current function at ≥50° . The same format is used for all the equivalent current plots, except that the spacing between contours will be 30 kA instead of the 20 kA used in this figure.

where $P_n^m$ is the Schmidt seminormalized Legendre function of degree $n$ and order $m$. $a$ is the Earth radius. We only use terms with $n \leq 30$, $m \leq 10$, $n \geq m$, and $n-m$ odd. The last requirement, also used by *Friis-Christensen et al.* [1984], makes $V$ antisymmetrical about the equator (we use data only from the Northern Hemisphere). The radius $r$ was set equal to $a$ for all magnetometers. Two sets of the coefficients $A_n^m$ and $B_n^m$ are estimated using weighted least squares. One set of coefficients represents the potential associated with external sources and the other with internal sources. The latter can be interpreted as currents induced in the ground by the ionospheric currents. The separation of external and internal sources is made possible by use of the vertical magnetic field component. The weight for each equation is defined as the multiplicative inverse of the standard error of the mean vector component in each grid cell, unless the standard error is $\leq$1 nT, in which case the weight is 1.

The longitude $\phi_{\text{MLT}}$ in equation (3) is the magnetic local time, in radians

$$\phi_{\text{MLT}} = \phi_q - \phi_{\text{noon}} + \pi. \tag{4}$$

We define magnetic noon, $\phi_{\text{noon}}$, as the QD meridian that maps to the subsolar point at an Earth-centered sphere with radius $\gg 1R_E$. A large radius is chosen in order to avoid the influence of nondipole magnetic field structures on determining which magnetic longitude is most strongly facing the Sun. A standard way of computing the noon meridian is to use the subsolar point on ground, in which case features such as the South Atlantic anomaly can affect the result. Since we focus on high latitudes, which map to large radii, it is reasonable to assume that the dipole component of the field is more relevant in this regard.

With the coefficients for the external potential determined, an equivalent current function $\Psi$ can be calculated as [e.g., *Pothier et al.*, 2015]

$$\Psi = \frac{a}{\mu_0}\sum_{n,m}\frac{2n+1}{n+1}\left(\frac{a+h}{a}\right)^n P_n^m(\lambda_q)\left[A_{n,e}^m\cos(m\phi_{\text{MLT}}) + B_{n,e}^m\sin(m\phi_{\text{MLT}})\right]. \tag{5}$$

It relates to the vector equivalent current as $\mathbf{j} = \hat{\mathbf{r}} \times \nabla\Psi$, where $\hat{\mathbf{r}}$ is a unit vector in the radial direction. $h$ is the height of the equivalent current, chosen here to be 110 km. A contour plot of $\Psi$ for all the SuperMAG data used in this study is shown in Figure 1c. The contour spacing is 20 kA flows between each contour. Negative contours indicate lower values of $\Psi$ than solid contours. We use the same format throughout the paper, but the MLT labels are omitted. We also use a contour spacing of 30 kA in the next section, since the solar wind driving and thus the currents are stronger. The number of vectors used to make the average vectors is indicated in the top left corner of the plot. The lower left corner shows the misfit between the modeled field and the average vector components, quantified as the weighted root-mean-square difference. The number at the lower right shows the total current flowing between the maximum and minimum of $\Psi$ in the $\geq$50° QD latitude region.

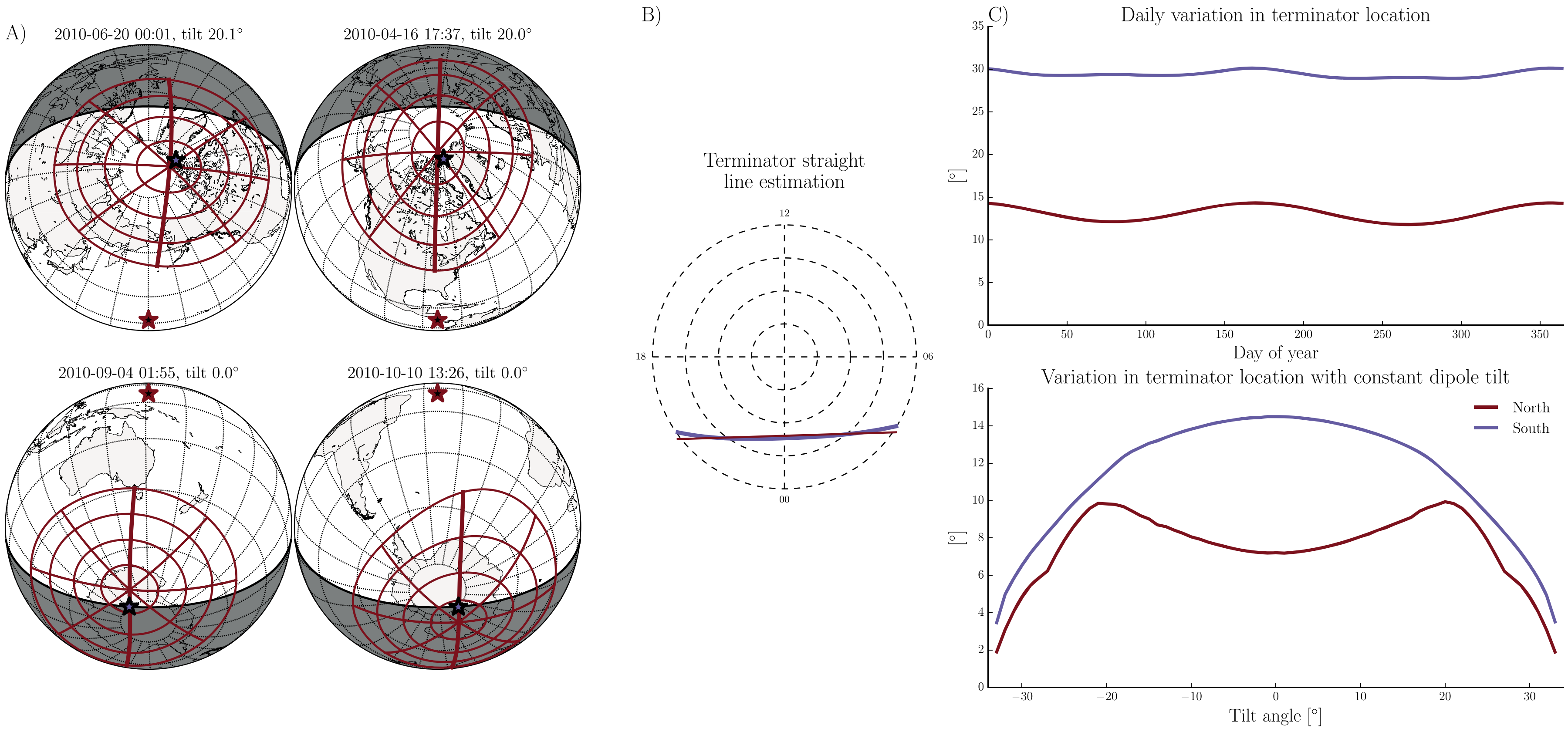

**Figure 2.** (a) Four examples of how the terminator crosses the noon-midnight meridian (bold red) at different locations with the same dipole tilt angle. The red stars mark the subsolar points, and the black stars mark the dipole pole locations. In the top row, the dipole tilt angle is close to 20° , but different fractions of the QD grid are sunlit. In the bottom row the same effect is shown for the Southern Hemisphere. (b) Illustration of how the sunlight terminator (blue) is fitted by straight line (red). We use the intersection with the noon-midnight meridian and the angle of the line as selection parameters in the statistics. (c) Daily variation in the terminator crossing point with the QD noon-midnight meridian (top) and the variation in this crossing point within 1° wide dipole tilt angle bins (bottom). Calculations are based on IGRF-12 coefficients for 2010.

### 2.2. Sunlight Terminator Location in Magnetic Coordinates

A key selection parameter for the results in this study is the position of the sunlight terminator. We define the sunlight terminator as the contour of $\chi = 90°$ , where $\chi$ is the solar zenith angle. On the sunward side of this contour the solar EUV-induced conductance increases in proportion to $\sqrt{\cos \chi}$, according to *Robinson and Vondrak* [1984]. On the nightside of this line, ionization predominantly stems from auroral particle precipitation. However, this is a simplification, since it would lead to an infinitely strong gradient at the terminator. In reality scattering of sunlight, along with antisunward plasma transport, smooths the conductance gradient at the terminator so that it is probably smaller than those associated with auroral precipitation. Since the location and intensity of the precipitation are less predictable than solar EUV flux, it is more convenient to use the sunlight terminator as a selection parameter when examining the effect of ionospheric conductance statistically.

Although it is more common to use the month or dipole tilt angle as selection parameter in studies of seasonal variations of ionospheric electrodynamics, we will argue that the terminator allows for a more precise determination of the sunlight distribution. This is illustrated in Figure 2a, which shows two examples from each hemispheres from periods when the dipole tilt angle was the same. A quasi-dipole grid is also shown (red) with the noon-midnight meridian in bold. In both examples the dipole tilt angle was approximately the same (20° for the Northern Hemisphere example and 0° for the Southern Hemisphere), but the terminator (black contour) crossed the noon-midnight QD meridian at different places. The distribution of sunlight on the QD grid is therefore significantly different in the two cases, although the dipole tilt angle is the same. This is due to the offset between the apex and dipole poles, and the distortion of the QD grid because of Earth's nondipolar field. Since the sunlight terminator is not a straight line in quasi-dipole coordinates, we choose to parametrize the $\chi = 90°$ contour by a linear fit. An example is shown in Figure 2b, with the terminator in blue and the fitted line in red. We use both the slope and the position (intersection of the terminator with the noon-midnight meridian) as selection parameters in the statistics presented in the next section.

Figure 2c shows the variability of the terminator line intersection with the noon-midnight meridian. The upper part shows the daily variation, in degrees, along the noon-midnight meridian in the Northern (red) and Southern (blue) Hemispheres. The variability is clearly larger in the south, due to the larger offset between the geographic and apex poles. The dipole tilt angle does not encompass this interhemispheric asymmetry.

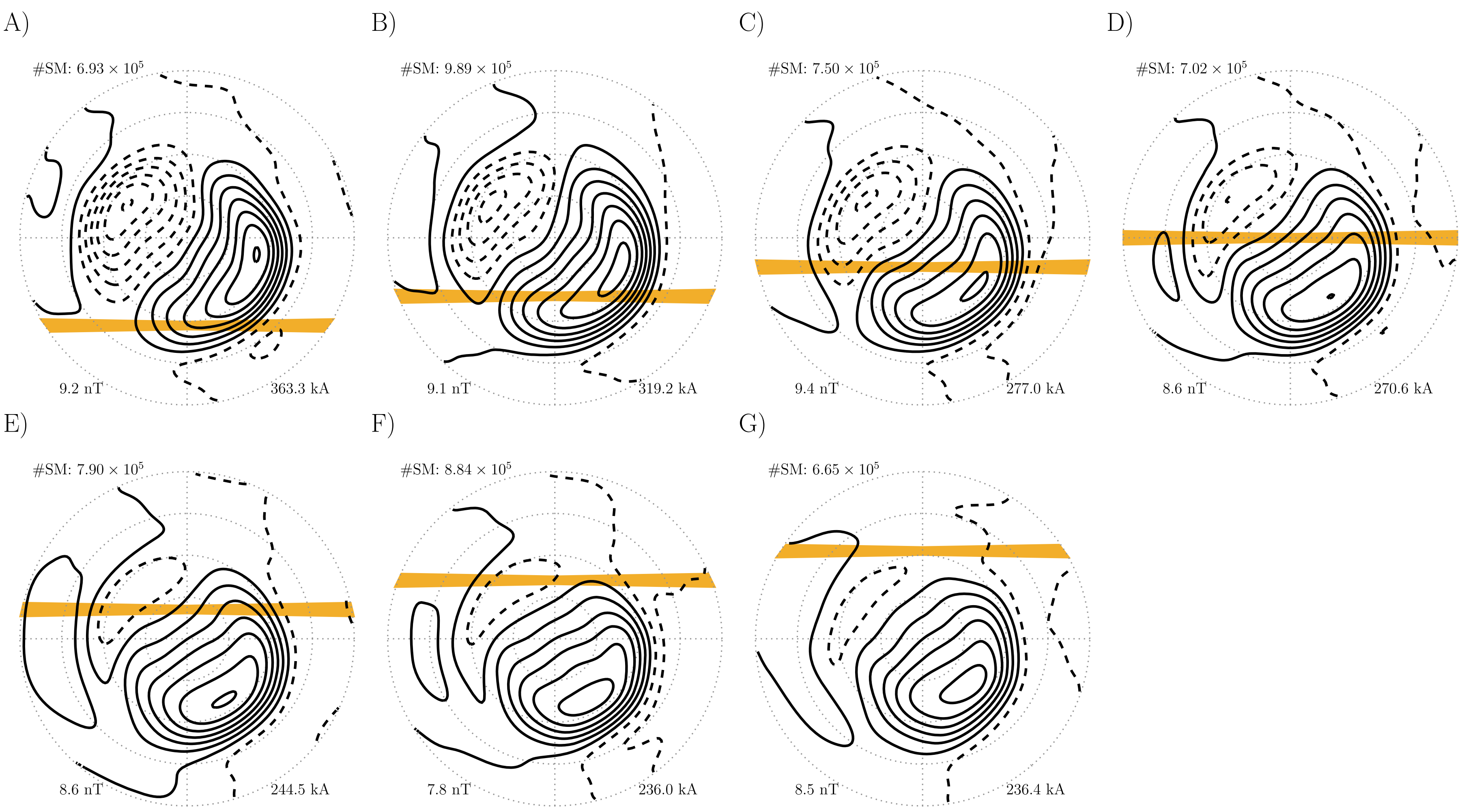

**Figure 3.** Equivalent currents with magnetometer data sorted according the position and orientation of the sunlight terminator, with IMF $B_z < -2$ nT. The colored strips show the limited range of locations of the terminator at the time when the data used to calculate the equivalent current were measured. The equivalent current is shown as black contours. The number of measurements used in the calculations is shown in the top left corners. The total current flowing between the maximum and minimum of the equivalent current function is shown in the bottom right. The plots extend to 50° QD latitude. Magnetic noon is on top, midnight on the bottom, dusk to the left and dawn to the right. The same format will be used for similar plots throughout the paper.

The lower part of the figure shows the variability in the terminator line's noon-midnight intersections within 1° wide dipole tilt angle bins. It shows, for example, that when the dipole tilt angle is between −0.5° and 0.5°, the noon-midnight intersection can vary by approximately 14° in the south and 7° in the north. In the Northern Hemisphere, the variability is greatest, up to ≈10°, when the dipole tilt angle is close to ≈±20°.

Since we use the terminator line instead of the dipole tilt angle as selection parameter, we expect differences between the sunlit and dark part of the ionosphere to appear more strongly in the present study compared to previous studies.

## 3. Observations

In this section we present equivalent current patterns based on magnetometer data that are selected according to the sunlight terminator location and IMF orientation. In all except for one figure, the IMF $B_z$ (GSM) is less than −2 nT, which ensures that on average there will be significant energy transfer from the solar wind to the magnetosphere.

Figure 3 shows equivalent current patterns that correspond to seven different locations of the sunlight terminator. The colored strips show the limited range of locations of the fitted terminator line at the time when the data used to calculate the equivalent current were measured. The equivalent current is shown as black contours. At the top left in each panel the number of data points used in the calculations is shown. The total current flowing between the maximum and minimum of the current function is shown on the bottom right in each panel.

Figure 3a shows that when the polar region is predominantly sunlit, the equivalent current is largely a two-cell pattern. As it becomes increasingly dark, the dusk cell diminishes and, eventually, almost disappears. The dawn cell simultaneously rotates clockwise, so that the current across the polar cap gets a significant dawnward component. We also see a clear tendency that the total current across the polar cap decreases as the terminator moves toward the dayside, consistent with reduced conductivity from sunlight.

**Table 1.** Average Values for Solar Wind Flow Speed, IMF $B_y$, IMF $B_z$ (GSM Coordinates), and Proton Density, Based On OMNI Data From 1981 to 2014, Binned by the Same Selection Criteria as the Panels in Figure 3, i.e., by the Location of the Sunlight Terminator in the Northern Hemisphere

| Figure 3 | Flow Speed (km/s) | $B_z$ (nT) | $B_y$ (nT) | Proton Density (cm$^{-3}$) |
|---|---|---|---|---|
| Figure 3a | 452.2 | −4.1 | 0.1 | 6.6 |
| Figure 3b | 442.2 | −4.1 | 0.1 | 6.9 |
| Figure 3c | 438.2 | −4.2 | 0.3 | 7.7 |
| Figure 3d | 442.8 | −4.0 | −0.1 | 7.0 |
| Figure 3e | 437.4 | −4.2 | 0.0 | 7.2 |
| Figure 3f | 436.3 | −3.9 | 0.5 | 7.3 |
| Figure 3g | 442.6 | −4.5 | 0.1 | 7.7 |

The systematic changes between the panels in Figure 3 are not an effect of a bias in solar wind parameters, since the data in the different panels only differ by the location of the sunlight terminator, which is not correlated with the solar wind and IMF. The average solar wind parameters for the same selection criteria as in Figures 3a–3g are presented in Table 1.

In Figure 3 the angle of the terminator line with the dawn-dusk meridian is at most 1°. The terminator line can, however, intersect the noon-midnight meridian at an angle. The effect on the equivalent current is shown in Figure 4, where the noon-midnight intersection is held fixed, but the angle is varied. We see that when the duskside is more sunlit, the total current seems to increase slightly.

### 3.1. Effects of $F_{10.7}$

The effects seen in Figure 3 indicate that sunlight strongly affects the magnitude and morphology of the equivalent currents. To investigate this further, we examine the effects of variations in solar EUV flux, as parametrized by the 10.7 cm wavelength radio flux ($F_{10.7}$). *Moen and Brekke* [1993] showed that the Hall and Pedersen conductances change approximately as the square root of $F_{10.7}$. Figure 5 shows four equivalent current patterns based on data which are selected according to different $F_{10.7}$ values. The $F_{10.7}$ bin limits represent the quartiles during the 1981–2014 period, which covers three solar cycles. Since $F_{10.7}$ varies with the solar cycle, it has a certain correlation with solar wind velocity and magnetic field values. To reduce the effect of variations in these driving parameters, we also require the GSM $y$ component of the convection electric field in the solar wind, i.e., $E_y = v_x B_z$, to be between 1 and 2 mV/m. This constraint is not necessary in Figures 3 and 4, since the selection criteria there are not correlated with the solar wind (Table 1).

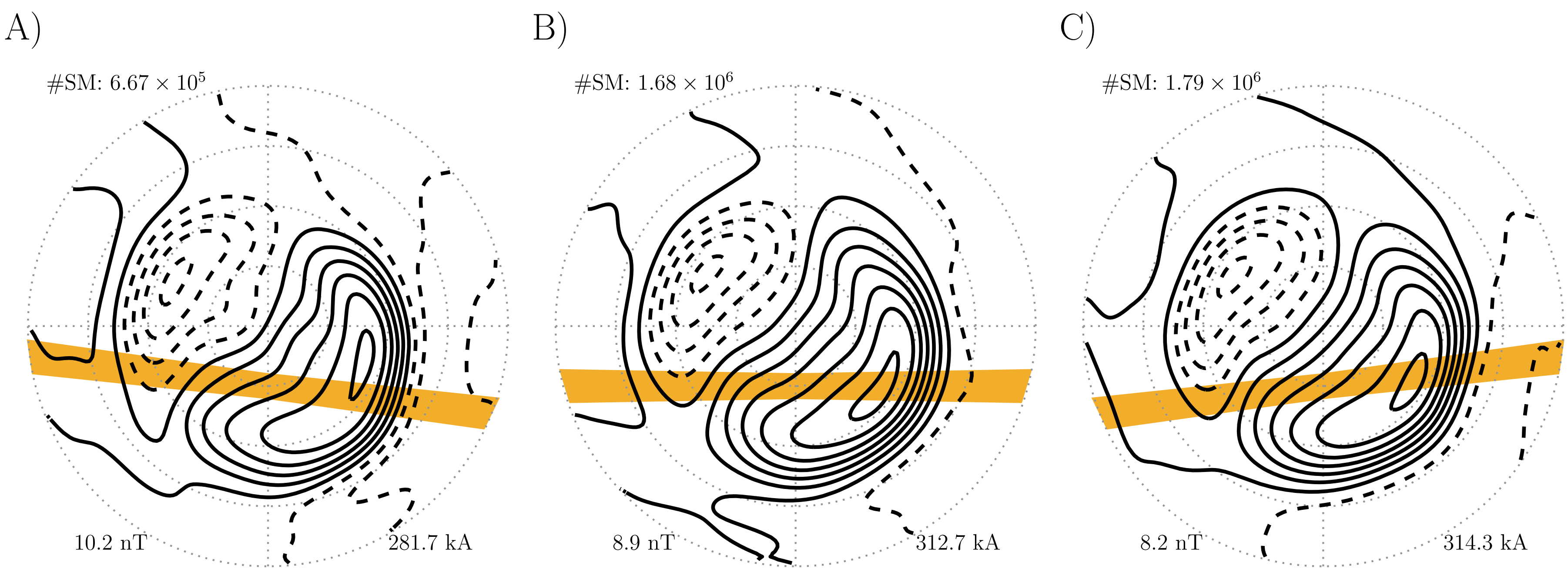

**Figure 4.** Equivalent currents for periods when the IMF $B_z$ was less than −2 nT and different orientations of the terminator line. The terminator intersects the noon-midnight meridian at 80° ± 2° on the nightside in all plots. The angle of the terminator line is 0° ± 1° in the middle plot and ±7° and ±1° at left and right. The format is otherwise the same as in Figure 3.

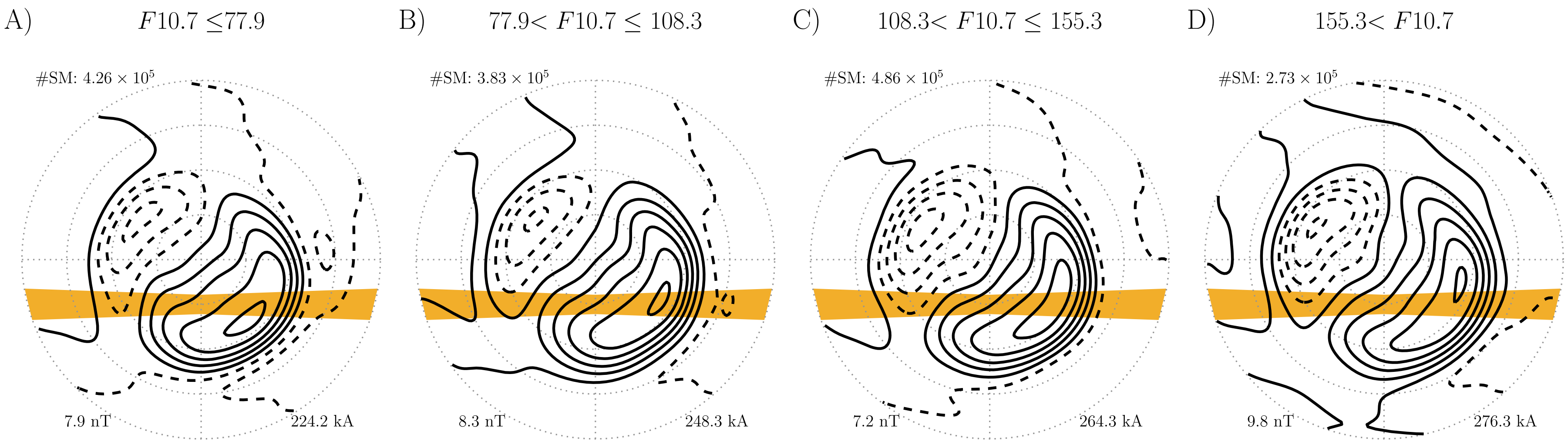

**Figure 5.** Equivalent currents based on magnetometer measurements sorted according the value of $F_{10.7}$, indicated above each plot. The solar wind convection electric field GSM $y$ component was held fixed at values between 1 and 2 mV/m. Otherwise, the format is the same as in Figure 3.

Figure 5 shows that on the nightside of the terminator, the equivalent current patterns are very similar. On the dayside, however, the current cell at dusk clearly increases with increasing $F_{10.7}$. We also see a tendency that the overall current increases with increasing $F_{10.7}$. Further, as $F_{10.7}$ increases, the sunlit part of the current pattern gets more aligned with the noon-midnight meridian.

These patterns thus show that $F_{10.7}$ affects not only the magnitude of the current but also its morphology. The contrast between the dayside pattern and nightside pattern increases with increasing $F_{10.7}$, which is a clear indication that the differences that we see across the terminator are due to sunlight and not a magnetospheric effect associated with the tilt angle.

### 3.2. Effects of Rotation of IMF $B_y$

The equivalent currents in the sunlit ionosphere in Figures 3 and 5 largely resemble statistical patterns of ionospheric convection for IMF $B_z$ negative conditions [*Heppner and Maynard*, 1987; *Weimer*, 2005; *Haaland et al.*, 2007; *Pettigrew et al.*, 2010]. Such patterns are known to change in response to changes in the IMF $B_y$ component. The corresponding change in equivalent currents is shown in Figure 6. All data points that were used to make this figure were measured when the IMF $B_z$ was less than −2 nT. Figures 6a, 6d, and 6g are based on measurements when the IMF $B_y$ was more strongly negative than $B_z$. Figures 6b, 6e, and 6h are based on measurements from times when $|B_y|$ is less than $|B_z|$ (neutral $B_y$). Figures 6c, 6f, and 6i correspond to $B_y$ positive conditions, $B_y$ being larger than $|B_z|$. As in Figures 3, 4, and 5, the colored strips represent the location of the sunlight terminator.

We see that in Figures 6a–6c, which represent sunlit conditions, the equivalent current flows in primarily two cells, which are different depending on the sign of IMF $B_y$. The $B_y$ influence is similar to what is seen in the statistical convection papers in the studies cited above: For $B_y$ negative, the dawn cell is relatively circular, and the dusk cell is more crescent shaped. The situation is reversed for $B_y$ positive conditions, and the crescent/round cells are even more clear. When $B_y$ is neutral, the cells are largely symmetrical on the dayside. The similarities between equivalent currents and statistical convection patterns strongly suggest that the Hall current system dominates the equivalent currents in sunlit conditions.

In dark conditions, the dusk cell is almost absent for all signs of $B_y$. The dominating dawn cell rotates only slightly clockwise as $B_y$ changes from negative to positive. More striking is the difference in magnitude, which is approximately 50%. An interpretation of this asymmetry is given in the next section.

## 4. Discussion

We have presented equivalent current functions based on an, to our knowledge, unprecedentedly large set of magnetometer data, spanning three solar cycles. We have particularly investigated the effect of variations in sunlight exposure on the ionosphere during relatively strong solar wind driving. The sunlight exposure has been parametrized in terms of the sunlight terminator position in magnetic quasi-dipole coordinates. The use of quasi-dipole coordinates leads to better accuracy compared to orthogonal magnetic coordinates, as the effect of longitudinal variations in the Earth's magnetic field is reduced [*Gasda and Richmond*, 1998; *Laundal and Gjerloev*, 2014]. A longitudinal bias would otherwise be present in the equivalent current patterns presented here, since certain orientations and locations of the terminator occur primarily at specific universal times. This UT bias implies that the same magnetometers largely cover the same MLTs in the statistics.

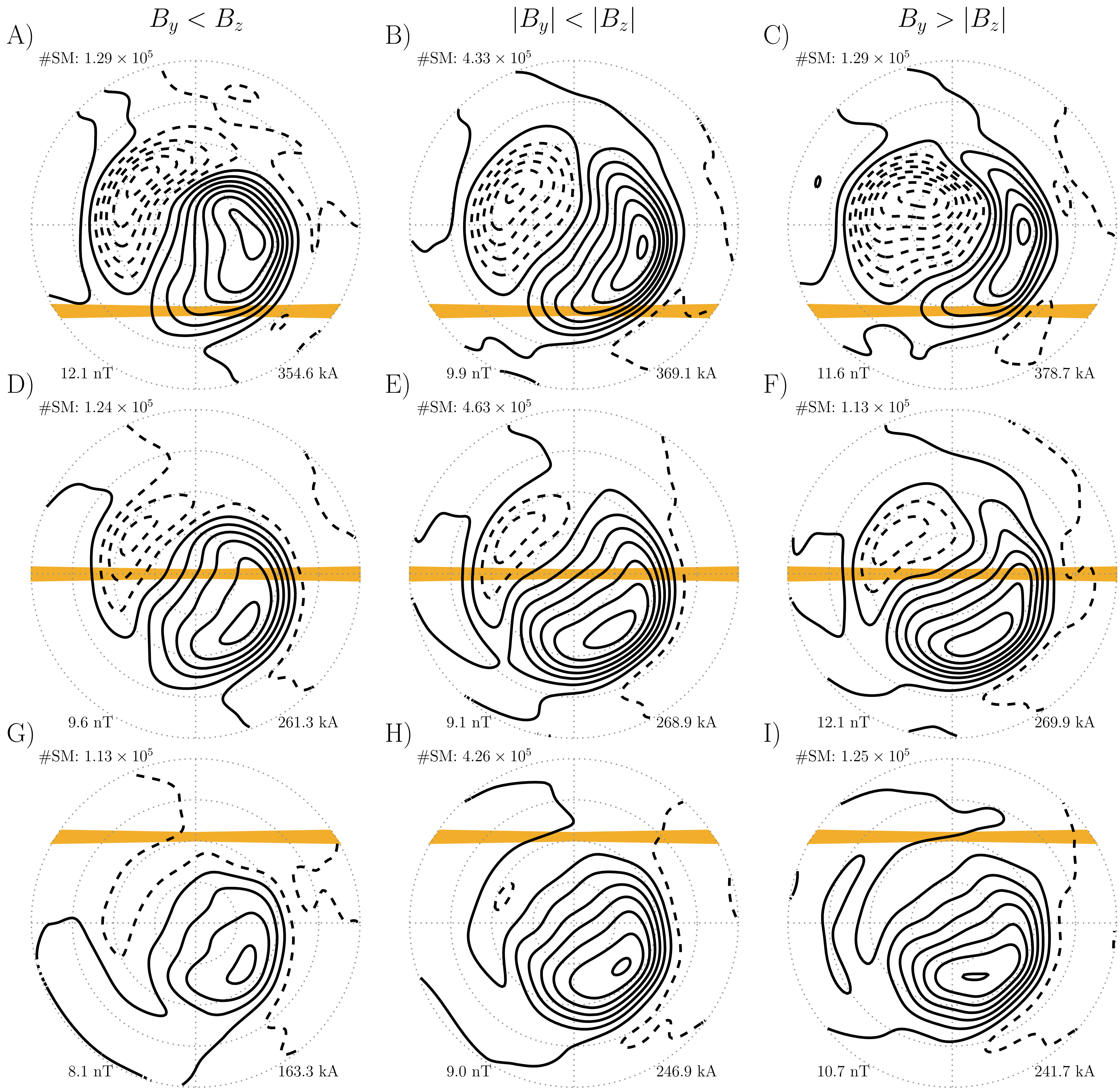

**Figure 6.** Equivalent currents with magnetometer data sorted according the position and orientation of the sunlight terminator, with IMF $B_z < -2$ nT. The value of IMF $B_y$ is indicated above each column: (a, d, and g) $B_y$ negative conditions, (b, e, and h) $B_y$ neutral, and (c, f, and i) $B_y$ positive conditions. Otherwise, the format is the same as in Figure 3.

The presented patterns show a very systematic behavior with changes in the sunlight terminator location, which strongly suggests that the effects shown in this paper are real and significant. The root-mean-square misfit of the fitted potentials with respect to the average vectors is also quite small, close to 10 nT in all figures presented in section 3 (the misfit is given in the lower left corners of each panel). This shows that the spherical harmonic expansion is a good representation of the average vectors. The variability in the data which was used to calculate the average vectors is, however, significant. The variability of the vectors used in Figure 3d (with the terminator at the dawn-dusk meridian) is illustrated in Figure 7. Figure 7a shows the standard deviations of the vector magnitudes. These maximize where the equivalent current (and field perturbations) is largest. The standard deviation is typically close to the mean vector magnitude. Figure 7b shows the Ψ parameter as defined by *Gjerloev and Hoffman* [2014] (not to be confused with the equivalent current function of equation (5)). This parameter is a measure of the variability in direction, independent of magnitude. $\Psi = 1$ means that all vectors are parallel, and Ψ close to zero means that there is a great deal of variability in direction. We see that the latter is true in most grid cells. There is a weak tendency that the vectors are more aligned in the auroral region, where the vector magnitudes are largest, but the pattern is not as clear as Figure 7a. Similar results have been found for the vectors used in the other panels of Figure 3 (not shown). Figure 7 shows that the average vectors and corresponding equivalent current patterns must be interpreted as exactly that average large-scale patterns. We cannot expect these to be a good representation of the magnetic field and currents at any given time due to the great variability in the system. In our analysis we do not differentiate between different states of the magnetosphere-ionosphere (M-I) system. We have organized our data using the solar zenith angle, IMF conditions, and $F_{10.7}$, but none of these allow a separation of, for example, substorm versus nonsubstorm conditions or storms versus quiet. These processes internal to the M-I system are highly variable in space and time and thus lead to a high degree of scatter in our data. The point of this paper, however, is to investigate the equivalent currents dependence on the listed external conditions. What we note is that despite the expected scatter, we find clear systematic dependencies of the equivalent current system on these conditions.

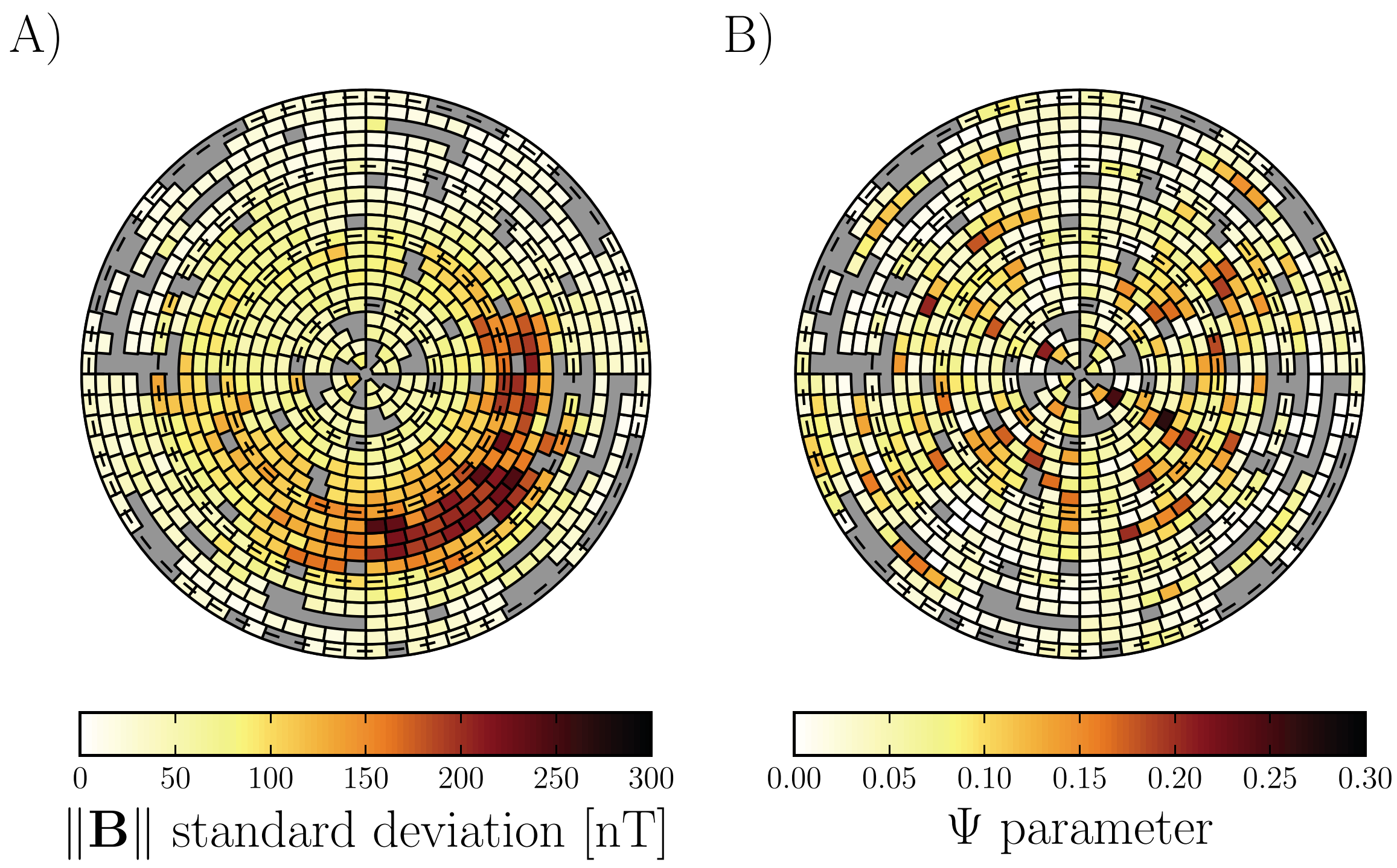

**Figure 7.** Illustration of the variability of the data used to calculate the average vectors used in Figure 3d. (a) Standard deviation of the magnetic field perturbation magnitudes within each bin. (b) The Ψ parameter [*Gjerloev and Hoffman*, 2014], measuring the stability in the vector directions. Only bins with > 100 vectors were used to produce this plot.

The most striking difference between the sunlit and dark part of the ionosphere is that the equivalent current patterns resemble two-cell opposite convection patterns on the sunlit side of the terminator but not on the dark side. On the dark side the dawn current cell strongly dominates, and the dusk cell is almost absent. The sunlit-dark difference is consistent with the study presented by *Friis-Christensen and Wilhjelm* [1975], who used eight magnetic observatories to produce global maps of equivalent current vectors for different seasons. The difference across the terminator is clearly controlled by the sunlight, and not some magnetospheric effect associated with the correlated change in dipole tilt angle. The basis for this conclusion is primarily the observations that the day-night contrast increases with solar EUV flux ($F_{10.7}$). There also seems to be a small variation in the patterns with respect to the terminator orientation, when its noon-midnight intersection is fixed.

*Laundal et al.* [2015] compared SuperMAG ground magnetic field vectors at polar cap latitudes to simultaneous and coincident convection electric field measurements. They found that in sunlight the equivalent currents tend to be antiparallel to convection and perpendicular in darkness. The similarity between our current patterns sunward of the terminator and statistical convection patterns strongly suggests that in sunlight the equivalent current is largely dominated by the Hall current component also at lower latitudes. The equivalent current on the sunlit side of the terminator rotates anticlockwise with increasing $F_{10.7}$ (Figure 5). This rotation is consistent with a stronger alignment with the expected Hall current system. We therefore interpret the rotation as an increasingly important Hall current contribution to the equivalent current when solar-induced ionospheric conductivity increases. The relative increase in Hall current contribution does not necessarily imply that Hall currents increase more than Birkeland currents with increasing $F_{10.7}$; it may be that the conductivity distribution changes so as to align the divergence-free (observed) current and the Hall current.

**Table 2.** Average Values of the *AL* Index in June and December for Different Signs of $B_y$[a]

| | December | June |
|---|---|---|
| $B_y > 2$ nT | −144.9 nT | −102.2 nT |
| $B_y < -2$ nT | −68.9 nT | −151.6 nT |

[a]Both *AL* and $B_y$ were obtained from the 1 min OMNI data set from the period February 1981 to July 2015.

By the same logic as in the previous paragraph, the equivalent currents on the dark side of the terminator are not dominated by Hall currents. Statistical studies of convection [e.g., *Pettigrew et al.*, 2010] do not show a similar dawn cell dominance as is seen in the equivalent current. *Laundal et al.* [2015] found that in the dark polar cap, the equivalent current is consistent with the actual current being close to zero, so that the equivalent current and the curl-free current component balance. The latter represents the Birkeland current closure current. Thus, the equivalent current in this region is determined by the surrounding Birkeland current system. The results in the present paper show that the equivalent current in the polar cap becomes increasingly dawnward in darkness, consistent with it being antiparallel to the curl-free currents connecting the Region 1 currents at dawn and dusk. However, equatorward of the polar cap, the conductance is on average nonzero due to auroral particle precipitation. Here we do not expect the curl-free and equivalent currents to balance, and thus, we expect the Hall currents to make a significant contribution to ground magnetic disturbances.

We have shown that when the auroral region is predominantly dark, the equivalent current, which primarily consists of a dawn cell, is approximately 50% stronger when $B_y$ is positive compared to negative conditions (an opposite asymmetry would be expected in the Southern Hemisphere). This asymmetry was also noted by *Friis-Christensen and Wilhjelm* [1975]. This suggests that a similar asymmetry with respect to $B_y$ exists in the *AL* index, since this index is a measure of the peak westward electrojet, the equatorward part of the dawn equivalent current cell. Average values of *AL* in June and December are summarized in Table 2. A factor of 2 difference is seen in December. In June the asymmetry is opposite, but not as pronounced. *AL* is often used as a proxy for substorm activity. Other metrics of substorm activity, such as total flux closure, do not show a similar asymmetry with respect to $B_y$ [e.g., *Laundal et al.*, 2010]. Our results therefore imply that care should be taken when using *AL* to quantify substorm activity during northern winters.

Figure 6 also shows that the equivalent current cell in darkness rotates slightly clockwise as $B_y$ changes from negative to positive values. This is consistent with the observations of field-aligned currents at northern winter made by *Green et al.* [2009], in combination with the interpretation that the equivalent current across the dark polar cap is antiparallel with the horizontal Birkeland current closure. The Birkeland current patterns presented by *Green et al.* [2009] show that the dawn R1 current has a clear maximum on the dayside when $B_y<0$ and is more uniform along the dawn flank when $B_y>0$. The dusk R1 current maximizes on the nightside for both signs of $B_y$. Thus, when $B_y < 0$, the curl-free currents will be more aligned with the noon-midnight meridian than for positive $B_y$, consistent with the observed equivalent currents.

Figure 6 shows that during sunlit conditions, there is an azimuthal current around noon with opposite polarity for different signs of IMF $B_y$. This is known as the Svalgaard-Mansurov effect. It was first explained by *Jørgensen et al.* [1972] in terms of Hall currents flowing opposite to ionospheric convection: Ionospheric convection on the dayside will be forced azimuthally when there is a significant GSM *y* component in the IMF, due to the geometry of newly reconnected field lines [*Cowley*, 1981]. Figure 6 shows that this effect is only seen in the equivalent current when the dayside is sunlit. In darkness the azimuthal component of the current is toward dawn for both signs of $B_y$. We therefore conclude that the Svalgaard-Mansurov effect on average is only present in sunlight.

## 5. Conclusions

We have calculated and analyzed equivalent current patterns based on ground magnetometer measurements for different sunlight conditions in the ionosphere. The patterns resemble convection patterns on the sunlit side of the terminator but not on the dark side. Based on this observation, we conclude that (1) the global equivalent currents are dominated by Hall currents in sunlight and (2) in darkness there is a significant indirect contribution to the equivalent currents from the Birkeland currents. The Svalgaard-Mansurov effect is on average only present in sunlight. These results agree with the conclusions of *Laundal et al.* [2015] but show more clearly that the morphology of the global equivalent current is different at the dark and sunlit side of the terminator. The sunlit-dark contrast increases with $F_{10.7}$, which confirms that the observed effects depend on the solar EUV flux.

**Acknowledgments**
For the ground magnetometer data we gratefully acknowledge the following: Intermagnet; USGS, Jeffrey J. Love; CARISMA, PI Ian Mann; CANMOS; The S-RAMP Database, PI K. Yumoto and K. Shiokawa; The SPIDR database; AARI, PI Oleg Troshichev; The MACCS program, PI M. Engebretson; Geomagnetism Unit of the Geological Survey of Canada; GIMA; MEASURE, UCLA IGPP and Florida Institute of Technology; SAMBA, PI Eftyhia Zesta; 210 Chain, PI K. Yumoto; SAMNET, PI Farideh Honary; The institutes who maintain the IMAGE magnetometer array, PI Eija Tanskanen; PENGUIN; AUTUMN; PI Martin Connors; DTU Space, PI Juergen Matzka; South Pole and McMurdo Magnetometer, PI Louis J. Lanzarotti and PI Alan T. Weatherwax; ICESTAR; RAPIDMAG; PENGUIN; British Antarctic Survey; McMac, PI Peter Chi; BGS, PI Susan Macmillan; Pushkov Institute of Terrestrial Magnetism, Ionosphere and Radio Wave Propagation (IZMIRAN); GFZ, PI Juergen Matzka; MFGI, PI B. Heilig; IGFPAS, PI J. Reda; University of L'Aquila, PI M. Vellante; and SuperMAG, PI Jesper W. Gjerloev. The SuperMAG website, from which the data were obtained, can be found at http://supermag.jhuapl.edu/. The IMF, solar wind, and magnetic index data were provided through OMNIWeb by the Space Physics Data Facility(SPDF) and downloaded from ftp://spdf.gsfc.nasa.gov/pub/data/omni/high_res_omni/. The $F_{10.7}$ index was downloaded from http://lasp.colorado.edu/lisird/tss/noaa_radio_flux.html. This study was supported by the Research Council of Norway/CoE under contract 223252/F50.